\documentstyle[editedvolume,epsfig,numreferences]{crckapb}

\newcommand{\be}{\begin{equation}}
\newcommand{\ee}{\end{equation}}
\newcommand{\bea}{\begin{eqnarray}}
\newcommand{\eea}{\end{eqnarray}}

\newcommand{\beet}{\begin{equation*}}
\newcommand{\eeet}{\end{equation*}}
\newcommand{\beaet}{\begin{eqnarray*}}
\newcommand{\eeaet}{\end{eqnarray*}}
\newcommand{\bfig}{\begin{figure}}
\newcommand{\efig}{\end{figure}}
\newcommand{\bc}{\begin{center}}
\newcommand{\ec}{\end{center}}

\newcommand{\szz}{\sigma_{tt}}
\newcommand{\sxx}{\sigma_{xx}}
\newcommand{\szx}{\sigma_{tx}}
\newcommand{\sxz}{\sigma_{xt}}

\begin{opening}
\title{Models of stress propagation in granular media}

\author{J.P. Bouchaud}\author{P. Claudin}
\institute{Service de Physique de l'Etat Condens\'e,\\ CEA, Ormes des
Merisiers,\\ 91191 Gif-sur-Yvette, Cedex France. }
\author{M. E. Cates}\author{J. P. Wittmer}
\institute{Dept. of Physics and Astronomy, University of Edinburgh\\ JCMB
King's Buildings, Mayfield Road,\\
Edinburgh EH9 3JZ, UK. }

\end{opening}

\runningtitle{Models of stress propagation}

\begin{document}

\begin{abstract}
Stress patterns in static granular media exhibit unusual features when
compared to either liquids or elastic solids. Qualitatively, we attribute
these features to the presence of `stress paths', whose geometry depends on
the construction history and controls the propagation of stresses. Stress
paths can cause random focussing of stresses (large fluctuations) as well
as systematic
deflections (arching). We describe simple physical models that capture some
of these effects. In these models, the `stress paths' become identified
with the characteristic `light rays' of wavelike (hyperbolic) equations for
force propagation. Such models account for the  `pressure dip'  below
conical sandpiles built by pouring from a point source, and explain
qualitatively the large stress fluctuations  observed experimentally in
granular matter. The differences between this approach and more
conventional modelling strategies (based on elastoplastic or rigid-plastic
models) are highlighted, focusing on the role of boundary conditions. Our
models provide a continuum picture in which granular materials are viewed
as {\it fragile matter}, able to support without rearranging only a subset
of the static external loadings admissible for a normal elastic solid.
\end{abstract}

\section{Introduction}

Stress patterns in granular media exhibit some rather unusual features when
compared to either liquids or elastic solids. For example, the
vertical pressure below conical sandpiles does not follow the height of
material above a particular point, but rather has a {\it minimum} underneath
the apex of the pile \cite{Smid,Huntley,WCCB}. Furthermore, local stress
fluctuations are large, sometimes on length scales much larger than the
grain size. For
example, repeatedly pouring the very same amount of powder in a silo results in
fluctuations of the weight supported by the bottom plate of $20 \%$ or more
\cite{Brown,Clement}. Qualitatively, these features are attributed to the
presence
of {\it stress paths} which can focus the stress field into localized
regions and
also deflect it to cause ``arching" (see also \cite{Radjai,Baxter}, and
\cite{Dantuetc} for early qualitative experiments).

More quantitative experiments were recently performed by Liu et al. \cite{Liu},
Brockbank et al. \cite{Huntley} and Mueth et al. \cite{Nagel}, where the local
fluctuations of the normal stress deep inside a silo or at the base of a
sandpile
were measured. It was found that the stress probability distribution is rather
broad, decaying exponentially for large stresses. A simple `scalar' (one
component)
model for stress propagation was introduced and studied in detail
\cite{Liu,Copper}, which
predicts a stress probability distribution in good agreement with experimental
(and numerical) data. However, this model only considers the {\it vertical}
normal
component of the stress tensor, and discards all the other components.

A fully `tensorial' model for stress propagation in homogeneous granular
media was
proposed in \cite{BCC,WCCB,FPA} to account for the pressure `dip' described
above. The most striking feature of this model is that the stress propagation
equation is (at least in two dimensions) a {\it wave equation}, with the
vertical axis
playing the r\^ole of
time. In this model, `stress paths' naturally appear as
the characteristics -- or the `light rays' -- of the corresponding
{\it hyperbolic} equation. Note that the standard equations of elasticity are
{\it elliptic}; the fundamental difference between these two cases will be
discussed later. Both must also be contrasted with the scalar model, which
corresponds to a parabolic equation, and in which stresses travel almost
vertically.

The aim of this paper is to review some of the recent theoretical work in
this field. We will start by summarizing the content of the scalar `$q$-model',
which, although unsatisfactory in several respects, offers the advantage of
simplicity. Keeping the spirit of the `$q$-model', we then show that the
introduction of the shear stress fundamentally modifies the structure of the
equations and leads to wave-like propagation. Several issues concerning the
solution
of this wave equation are then discussed. We then consider some variants of
the wave equation, in particular to account for local inhomogeneities or for
anisotropy, for example induced by the construction history. 
(The importance of construction history in granular matter has been
acknowledged for at least a century \cite{Gudehus}.)
Among a family of models, the `Fixed Principal Axis' ({\sc fpa}) 
limit plays a special r\^ole which we
discuss in relation with experimental data on conical sandpiles. We then
consider
in more detail the differences between this approach and some more conventional
modelling strategies (based on elastoplastic or rigid-plastic models), and try
to shed some light on the controversy
that the ``hyperbolic" approach to sandpile modelling has raised
recently \cite{Savage,Cantelaube,Evesque}. This discussion focuses in
particular on
the r\^ole of {\it boundary conditions}.

Much of our discussion will, for clarity, be limited to two dimensional
piles. However, most of the recent experimental results are in three dimensions;
indeed it was in this context that the physical assumptions of
Refs.\cite{WCCB,FPA}
were made.

\section{The Scalar Model}
\label{Scalar}

The main assumption of the scalar model is that only the vertical normal
component
 of the stress tensor $w = \sigma_{zz}$ (the `weight') needs to be
considered. Supposing for simplicity that the grains reside on the nodes of a
two-dimensional lattice
(see figure \ref{liufig}), the simplest model for weight propagation is:
\be
\label{liudiscret}
w(i,j) = w_g + q_+(i-1,j-1)w(i-1,j-1) + q_-(i+1,j-1)w(i+1,j-1)
\ee
where `$w_g$' is the weight of a grain, and $q_\pm(i,j)$ are `transmission'
coefficients giving the fraction of weight which the grain $(i,j)$
transmits to its
right (resp. left) neighbour immediately below.
\bfig[hbt]
\bc
\epsfysize=6cm
\epsfbox{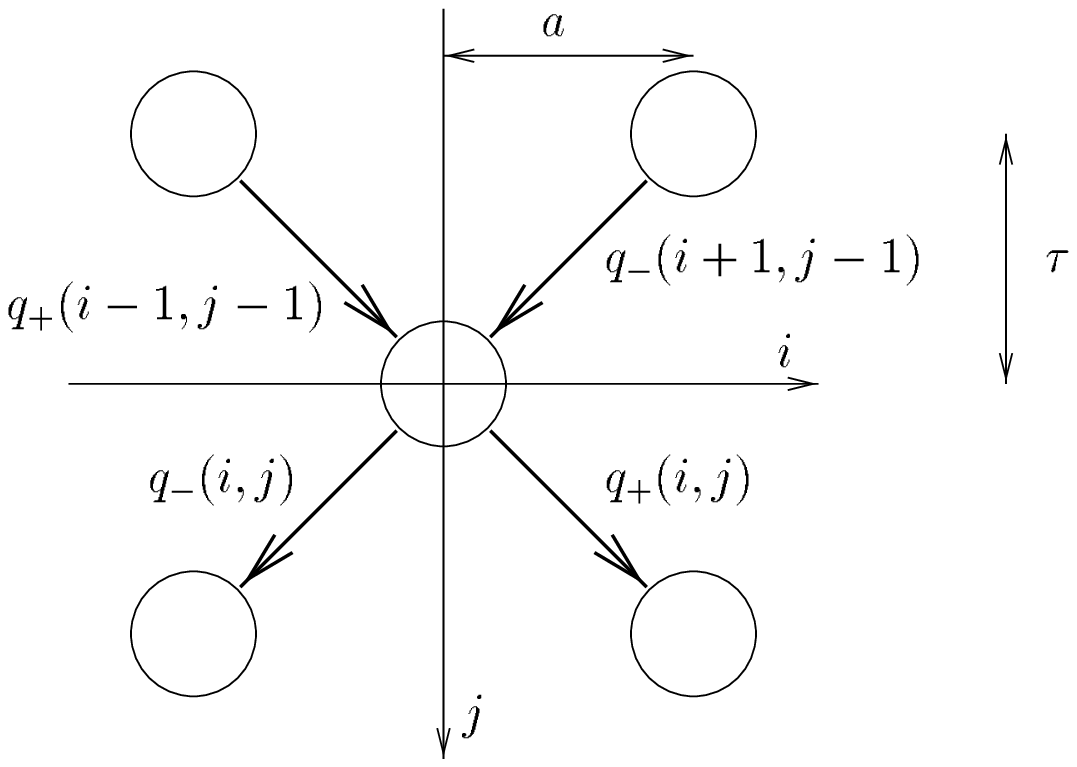}
\caption{The `q-model' model with two neighbours. $q_\pm$'s are
independent random variables, satisfying the weight conservation constraint:
$q_+(i,j) + q_-(i,j)=1$.
\label{liufig}}
\ec
\efig
Mass conservation imposes that $q_+(i,j) + q_-(i,j)=1$ for all $i,j$'s.
The case of an ordered pile of identical grains would correspond to
$q_\pm = \frac{1}{2}$. In this case, the equation (\ref{liudiscret}) describes
the `time'-evolution  (along the $j$ direction) of  the probability
density for a random walker on a line. The authors of \cite{Liu,Copper}
then propose to take into account the local disorder in packing, grain sizes and
shapes, etc., by choosing
$q_+(i,j)$ to be independent random numbers (subject to the above
constraint), for
example uniformly distributed  between $0$ and $1$. This case is
interesting because
it  leads to an exact solution for the local weight distribution $P(w)$, in
the limit
$j \to \infty$:
\be
P(w) = \frac{w}{W^2} \exp-\frac{w}{W}\label{exptail}
\ee
where $2 W= j w_g$ is the average weight. Liu et al. \cite{Liu,Copper} have
argued
that the exponential tail for large $w$ is generic; however, if the maximum
value
of $q$ is $q_M < 1$, one can show that $P(w)$ decays {\it faster} than an
exponential: \be \log P(w) \propto_{w \to \infty} -w^\beta \qquad
\beta=\frac{\log 2}{\log 2q_M}
\ee
(Notice that $\beta=1$ whenever $q_M=1$, and that $\beta = \infty$ in the
ordered case $q_M=1/2$). In this sense the exponential tail of $P^*(w)$ is not
universal: it requires the  possibility that one of the $q$ can be
arbitrarily close
to $1$, {\it i.e.} that there is a non zero probability that one grain is
entirely
bearing on one of its downward neighbours (local arching).

In the continuum limit, the $q$-model is equivalent to a diffusion equation with
a random convection term (related to $q_+-q_-$), for which several results are
known. However, on large scales, the random convection term merely
renormalises the
diffusion constant; hence, in this model, the response to a localised
overload at
the top of a pile spreads out diffusively as $\sqrt{D H}$, where $H$ is the
height
of the pile and $D$ is the diffusion constant, which is of the order of the
grain
size $a$. In the limit $H \gg a$, the spreading is negligible, which means
that the
weight essentially propagates vertically. Hence, the $q$-model predicts a `hump'
in the pressure profile underneath a sandpile, directly reflecting its shape.

A
way to accomodate a ``pressure dip" within this scalar picture was suggested by
Edwards (although in a slightly different language). Consider for example a
sandpile
built from a point source: the history of the grains will certainly inprint
a certain
oriented `texture' to the contact network, which can be modelled, within the
present scalar model, as a nonzero mean value of the `convection term', the
sign of
which depends on which side of the pile is chosen. Let us call $V_0$ the average
value of this  term on the $x\ge 0$ side of the pile, with $-V_0$ on the
other side. The differential equation describing propagation now reads, in the
absence of disorder:
\be \label{epe} \partial_t w + \partial_x \left [V_0 \
\mbox{sign}(x)w \right] = \rho +  D_0 \partial_{xx} w
\ee
Solving this equation in a sandpile geometry leads to a weight {\it
minimum} around
$x=0$.  Equation (\ref{epe}) gives a precise mathematical content to
Edwards' idea
of arching in sandpiles \cite{Edwards}, as the physical mechanism leading to a
`dip' in the pressure distribution \cite{Smid}. As discussed below (see also
\cite{WCCB,FPA}), this can be taken much further within a tensorial framework.
Let us also note that (\ref{epe}) can in fact be obtained
naturally within an extended $q$-model model, with an extra rule accounting
for the
fact that a grain can slide and lose contact with  one of its two downward
neighbours \cite{CB}. However, this extra sliding rule implicitly refers to the
existence of shear stresses, absent in the scalar model. It is more satisfactory
to recognize from the outset that stress has a tensorial, rather than
scalar, nature.
Models which do this are decribed next.

\section{A Tensorial Model}
\label{tensorial}

\subsection{The wave equation}

It is useful to start with a simple toy model for stress propagation, which
is the analogue
of the model presented in figure \ref{liufig}. We now consider the case of three
downward neighbours (see figure \ref{threeleg}), for a reason which will become
clear below. \bfig[hbt]
\bc
\epsfysize=4cm
\epsfbox{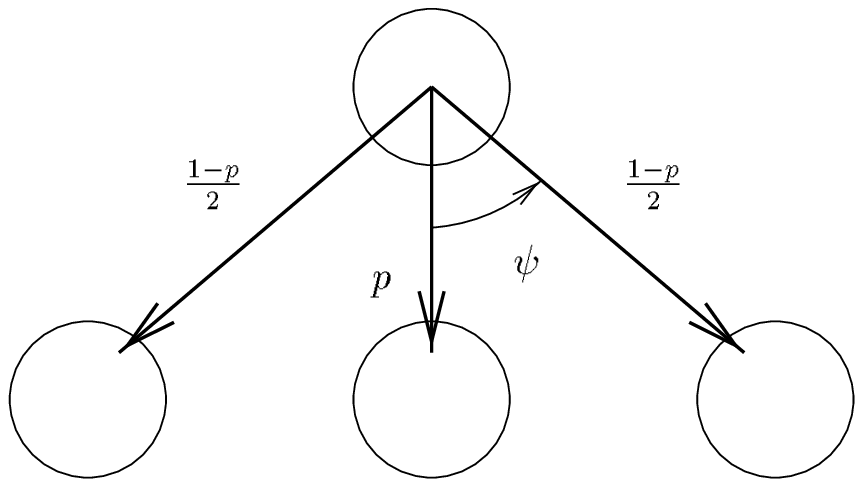}
\caption{Three neighbour configuration. Each grain tranmits two force components
to its downward neighbours. A fraction $p$ of the vertical component is
transmitted through the middle leg.
\label{threeleg}}
\ec
\efig
Each grain transmits to its downward neighbours not one, but two force
components:
one along
the vertical axis $t$ and one along $x$, which we call respectively $F_t(i,j)$
and $F_x(i,j)$. For simplicity, we assume that each `leg'
emerging from a given grain can only transport
the force parallel to itself (but more general rules could be invented).
Assuming
that the transmission rules are locally symmetric, and that a fraction $p
\leq 1$ of the
vertical component travels through the middle leg, we find:
\bea
F_x(i,j) & = & \frac{1}{2} \left[F_x(i-1,j-1)+F_x(i+1,j-1)\right] \nonumber \\
         &   & + \frac{1}{2} (1-p) \tan \psi
\left[F_t(i-1,j-1)-F_t(i+1,j-1)\right] \label{wemicro1}\\
F_t(i,j) & = & w_0 + p F_t(i,j-1) + \frac{1}{2}(1-p)
\left[F_t(i-1,j-1)+F_t(i+1,j-1)\right] \nonumber \\
         &   & +  \frac{1}{2 \tan \psi}
\left[F_x(i-1,j-1)-F_x(i+1,j-1)\right] \label{wemicro2}
\eea
where $\psi$ is the angle between grains, defined in figure \ref{threeleg}.
Taking
the continuum limit of the above equations leads to:
\bea
\label{equiF1}
\partial_t F_t + \partial_x F_x & = & \rho \\
\label{equiF2}
\partial_t F_x + \partial_x \left [ c_0^2 F_t \right ] & = & 0
\eea
where $c_0^2 \equiv (1-p) \tan^2 \psi$. Eliminating (say) $F_x$ between the
above
 two equations
leads to a {\it wave equation} for $F_t$, where
the vertical coordinate $t$ plays the r\^ole of time and
$c_0$ is the equivalent of the `speed of light'. In particular, the
stress does not propagate vertically, as it does in the scalar model, but rather
along two rays, each at a {\it non zero angle} $\pm\varphi$ such that
$c_0=\tan\varphi$. Note that
$\varphi
\neq \psi$ in general (unless $p=0$); the angle at which stress propagates has
nothing to do with  the underlying lattice structure and can take any value
depending on the local rules for force transmission. We chose
a three-leg model to illustrate this particular point.

The above derivation can be reformulated in terms of classical continuum
mechanics
as follows. Considering all stress tensor components $\sigma_{ij}$, the
equilibrium equation reads,
\bea
\label{equi1}
\partial_t \szz + \partial_x \sxz & = & \rho \\
\label{equi2}
\partial_t \szx + \partial_x \sxx & = & 0
\eea
Identifying the local average of $F_t$ with $\szz$ and that of $F_x$ with
$\szx$,
we see that the above equations (\ref{equiF1}, \ref{equiF2}) are actually
identical to
(\ref{equi1}, \ref{equi2}) provided $\szx=\sxz$ (which corresponds to the
absence of local torque)
and
\be
\sxx = c_0^2\, \szz \label{bcc}
\ee

This relation between normal stresses was postulated
in \cite{BCC} as the simplest ``constitutive relation" among stress components,
obeying the correct symmetries, that one can possibly assume.
The term ``constitutive relation" normally refers to a relation between
stress and strain, but the model under discussion has no strain variables
defined;
instead the particles are viewed as completely rigid. (Equations (\ref{equi1},
\ref{equi2}) are then indeterminate unless a further hypothesis relating
the stresses
themselves is made.) This particular choice can be interpreted as a {\it local}
Janssen approximation \cite{Janssen}.
We return later to a more detailed discussion of
closure equations of this type.
In the present case, the parameter $c_0^2$ must encode relevant details of
the local
geometry of the packing (friction, shape of grains, etc.) and may thereby
depend on
the {\it construction history} of the grain assembly. Only for simple,
`homogeneous'
histories (such as constructing a uniform sandbed using a sieve)  will
$c_0^2$ be
everywhere constant on the mesoscopic scale. Even then, unless an ordered
packing is
somehow created, local fluctuations of $c_0^2$ will always be present.

\subsection{Some simple situations}

The simplest situation is that of an infinitely wide layer of sand, of depth
$H$, with a localized ($\delta$-function) overload at the top. The weight at the
bottom then defines the {\it response function} of the wave equation,
which, in two
dimensions, is the sum of two $\delta$ peaks localised at $x=\pm c_0 H$.

Next, one
can consider the sandpile geometry. For a pile at repose, the position of
the free
surfaces are
$x =\pm c z$, where $c=\cot\phi$ with $\phi$ the repose angle. On these
surfaces,
all the stresses vanish. This boundary
condition is then (for given $c_0$ and $c$) sufficient to solve for the
stress field
everywhere in the pile. (See Section \ref{charz} below.) One then find that the
vertical normal component of the stress is piecewise linear as a function
of $x$. In
particular, for $-c_0 H
\leq x \leq c_0 H$, $\szz$ is {\it constant}. Therefore, in two dimensions, this
model \cite{BCC} predicts a flat-topped stress profile rather than a dip.

For a pile created by depositing grains from above (for example by sieving
sand onto
a disc) it is natural to expect the free surface to be a slip plane. (This
is a plane
across which the stress components saturate the Mohr-Coulomb condition.)
Interestingly, this provides a relation between $c_0$ and the friction angle
$\phi$, which reads: $c_0^2=1/(1+2\tan^2\phi)$ (note that since $c=1/\tan \phi$,
one has automatically $c > c_0$). Under these conditions one finds that the `plastic'
region (where the Mohr-Coulomb condition is saturated) extend inward from the
surface to encompass the outer `wings' of the pile (i.e. $c_0 z \leq |x| \leq
cz$); see figure \ref{comp}. This follows from the solution of the model
and is not
an {\it a priori} assumption, of the kind commonly made in elastoplastic
modelling
(e.g.,
\cite{Cantelaube}).  In three dimensions, a second closure relation is required
\cite{BCC}, but in all cases the stress profile has a broad maximum at the
center of
the pile. Now, however, the Mohr-Coulomb condition is only saturated in the
immediate vicinity of the free surface -- the `plastic' region has zero volume
in three dimensions \cite{BCC,FPA}.
\bfig[hbt]
\bc
\epsfysize=6cm
\epsfbox{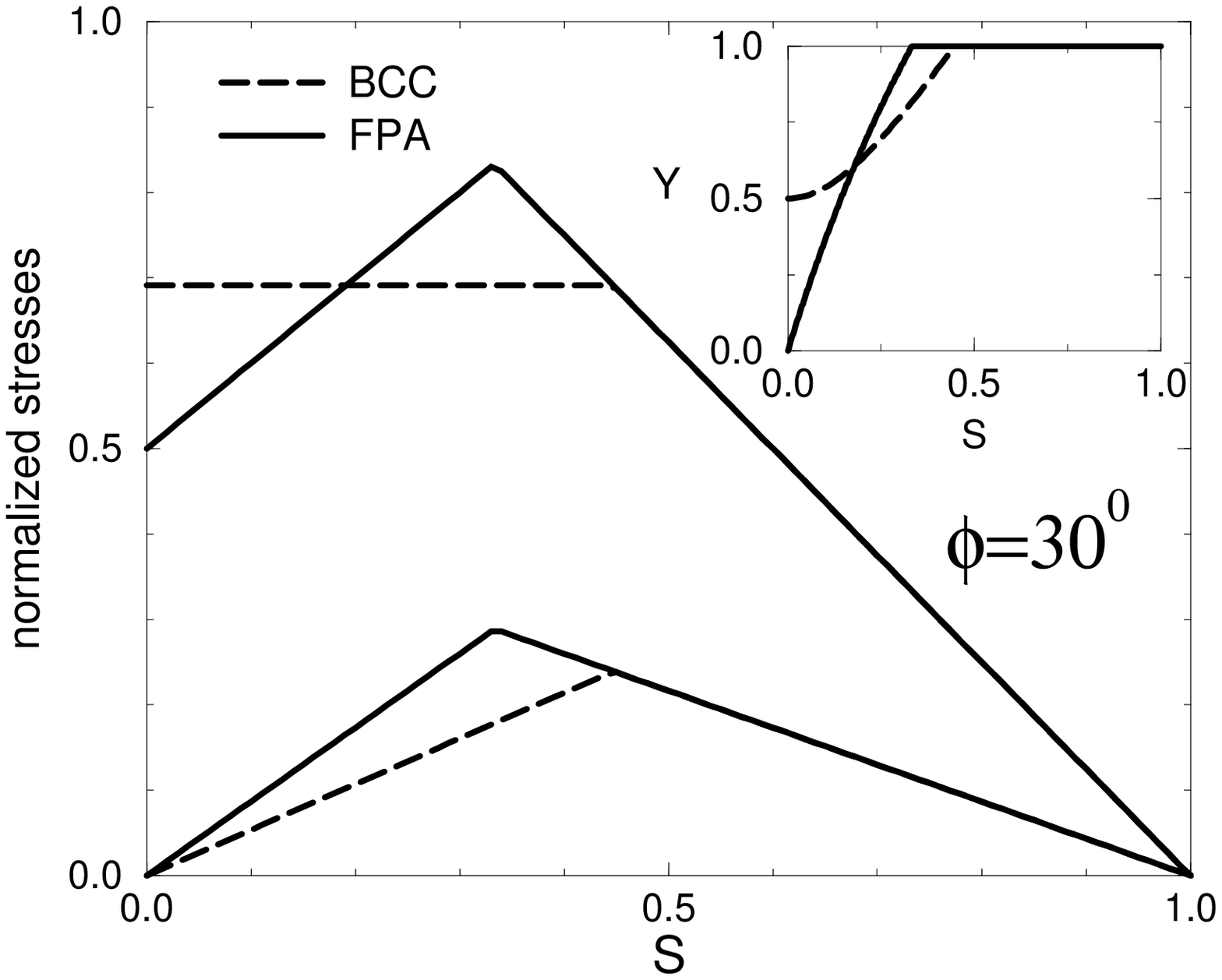}
\caption{Reduced normal (upper) and shear stress (lower) curves against
$S=r \tan\phi/t$ for the symmetric wave equation (``{\sc bcc}") and the
{\sc fpa}
model in two dimensions. Inset: the yield function $Y$. The Mohr-Coulomb
inequality
is saturated when
$Y=1$\label{comp}}
\ec
\efig

\section{Symmetries and Constitutive Relations}

Although above it was motivated in the context of a specific microscopic model,
the linear constitutive relation (\ref{bcc}) can be
viewed, independently of any microscopic model, as the
simplest closure equation compatible with the symmetries of the problem.
The latter
include a {\it local} reflection symmetry in which $x-x_0$ is changed to $x_0-x$
(with $x_0$ an arbitrary reflection plane) and also a form of ``dilational"
symmetry
known as RSF (``radial stress field") scaling. RSF scaling depends on the
absence of
any characteristic stress scale, which follows if the Young's modulus of
the grains
is sufficiently much larger than any stresses arising in the granular
assembly being
studied. Such scaling, which requires the stress distributions beneath piles of
different heights to have the same shape, is quite well confirmed in some
(but not all) experiments on conical sandpiles \cite{Smid,Huntley,Savage}.

Even with these two symmetries, one
can consider more complicated (nonlinear) constitutive relations among stresses,
which must be of the form
\cite{BCC}:
\be
\sxx = c_0^2 \szz {\cal F}\left(\frac{\sxz^2}{\szz^2}\right) \label{closgen}
\ee
Note that the Mohr-Coulomb condition itself can be written in this form.
Viewed as a
constitutive equation, it defines a rigid-plastic model whose physical
content is
to assume that, everywhere in the material, a plane can be found across
which slip
failure is about to occur. (The name ``incipient failure everywhere",
{\sc ife}, aptly describes this model \cite{BCC,WCCB,FPA}.)
All closures of the form (\ref{closgen}) lead to a hyperbolic equations for
stresses, although in the general case the characteristic directions of
propagation
(the `light rays' of the corresponding wave equation) depend on the loading and
therefore vary with position.

An interesting situation arises when local reflection symmetry is broken.
This is
the case, for example, in sandpiles created by pouring from a point source onto
a rough surface -- which is the usual mode of construction. In such a pile, all
grains arriving at the apex of the pile roll (in two dimensions) either to the
right or to the left. The two halves of the pile therefore have different
construction histories that are mirror images of each other. This violates local
reflection symmetry, and in general permits constitutive
equations such as:
\be
\sxx = c_0^2 \szz {\cal G}\left(\frac{{\rm{sign}}(x)\sxz}{\szz}\right)
\ee
The simplest case (found e.g. by expanding ${\cal G}$ to first order in the
shear to
normal stress ratio) corresponds to a family of (quasi-) linear constitutive
relations \cite{FPA}:
\be
\sxx = c_0^2 \szz + \mu\, {\rm{sign}}(x)\sxz \label{osl}
\ee
The previous, symmetrical, case has $\mu =0$. For nonzero $\mu$,
(\ref{osl}) again
leads to a wave equation, although this time {\it anisotropic}, in the
sense that
the two characteristic rays make asymmetric angles to the vertical axis. Note
that such a model can be obtained from rules such as those in figure
\ref{threeleg} simply by having an asymmetric partitioning of forces between
the supporting grains (or indeed by tilting the entire packing). Note also that
$x=0$ is a singular line across which the directions of propagation change
discontinuously. Microscopically,
$\mu \neq 0$ also leads to an unequal sharing of the weight of a grain
between the two characteristic rays propagating downward from it. For $\mu
< 0$, most
of the weight travel {\it outwards}; this provides, within a fully
tensorial model,
a mathematical description of the tendency to form arches, as developed by
Edwards
for the scalar case.

Solving these anisotropic wave equations for sandpiles in
two dimensions one again finds for
$\szz$ a piecewise linear function, which now has a sharp maximum at $x=0$
when $\mu > 0$, but a minimum for
$\mu < 0$, in accord with the arching scenario mentioned above (see figure
\ref{comp}). If one
furthermore imposes, as above, that the free surfaces are slip planes, one finds
a relation between $c_0^2$, $\mu$ and $\phi$.
\be
c_0^2=\frac{1}{1+2\tan^2\phi} \left [ 1-\mu\tan\phi \right ]
\ee
One again finds the result that the material throughout the outer wings
of the pile (exterior to the triangle formed by the characteristics passing
through the apex) are at incipient (Mohr-Coulomb) failure.

\subsection{The fpa Model}
Among possible forms of (\ref{osl}) there is a particular case, 
corresponding to $c_0^2=1$, which has some intriguing special properties:
\be
\sxx = \szz -2 {\rm{sign}}(x)\tan(\phi)\,\sxz \label{fpa}
\ee
(The resulting value of $\mu$ is that appropriate to the boundary
conditions stated above.) Specifically in this case, the  principal axis of the
stress tensor coincide with the `light rays' and hence have a fixed direction
throughout the pile (up to a reflection symmetry across the line
$x=0$). This particular case was called the {\sc fpa} (``fixed principal
axes") constitutive equation in \cite{WCCB,FPA}. It corresponds to an assumption
that the anisotropic texture of the medium imparts a local stress rule
which requires that the {\it orientation} of the stress tensor is fixed 
at the time of burial and, thereafter, cannot change under further loading 
(The values of the stress components themselves can, of course, change.). 
Since all material elements are buried near the
free surface of the pile, where the stress tensor orientation is known the fact
that the surface is itself a slip plane \cite{WCCB}, this fixes the
principal axes throughout the pile. The resulting stresses are shown in 
figure \ref{comp}.

At first sight, there is a problem with the {\sc fpa} description on the
symmetry axis of the pile ($x=0$), where the stress ellipsoid is required
simultaneously to have two conflicting orientations. However, since
$c_0^2=1$ the stress is isotropic at the centre, and there
is no conflict. However, this shows that the defining feature of the 
{\sc fpa} model is very easily lost; 
as soon as a slightly different $c_0^2$ is chosen, the principal
axes are no longer of fixed orientation but rotate smoothly as one passes
from one side of the pile to the other.

\section{Experiments on Cones and Wedges}
We do not have space here for an extended discussion of the experimental
data on cones and wedges; this can be found elsewhere \cite{usunpub}. 
Here we merely summarize some of the most important points.

\subsection{Cones}
The extension of the {\sc fpa} model to three dimensions, like the
other models discussed above, requires a second constitutive relation among
stresses to close the problem. Several well-known candidates \cite{footpod}
exist for this secondary closure and give similar results; 
these results are in surprisingly good agreement with the experimental data 
for the pressure dip in three
dimensional conical sandpiles built by pouring from a point source onto a rough
rigid support \cite{WCCB,FPA,Huntley}; see figure \ref{data}.
\bfig[hbt]
\bc
\epsfysize=6cm
\epsfbox{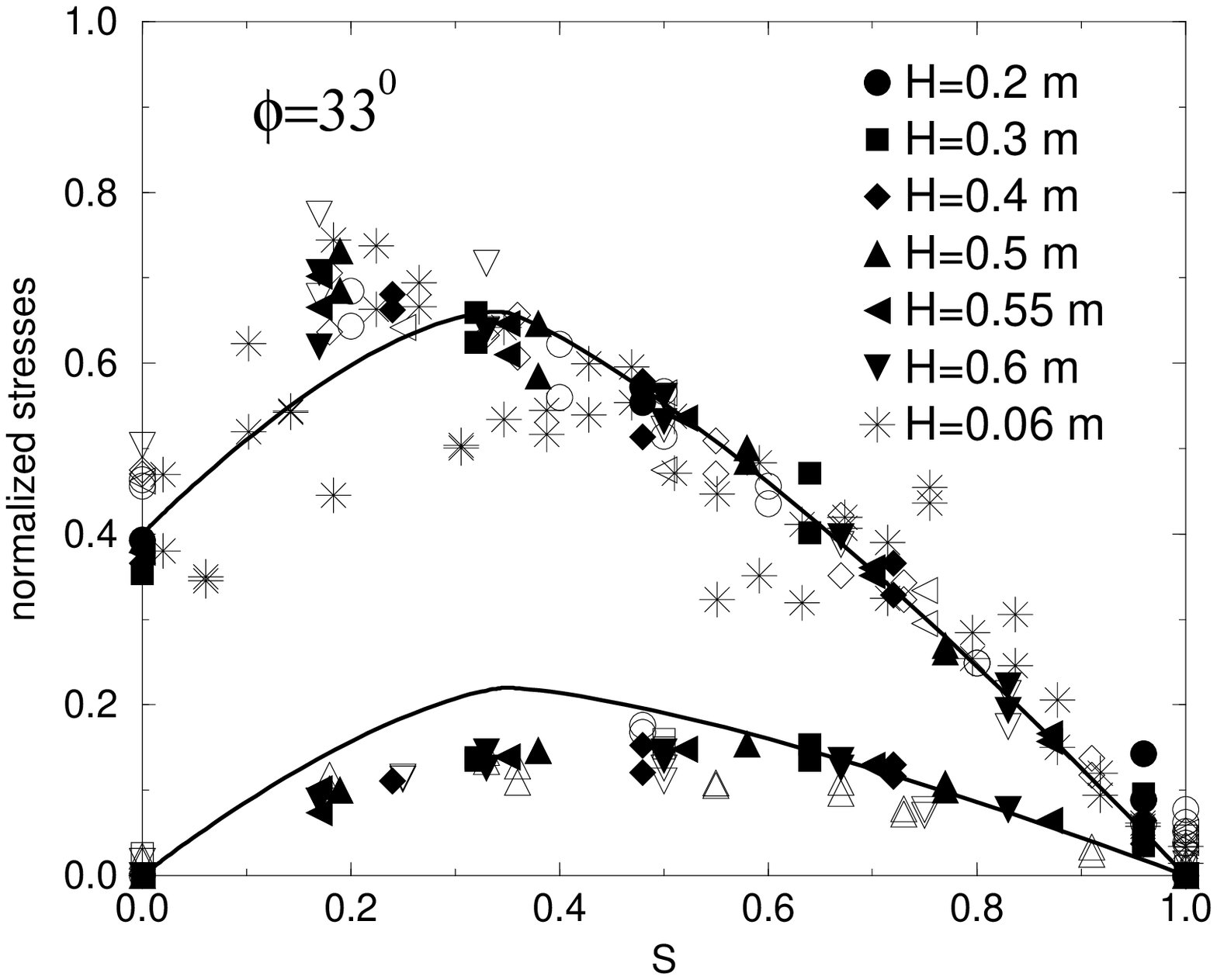}
\caption{Comparison of {\sc fpa} model (using a uniaxial secondary closure
\protect\cite{BCC,WCCB,FPA}) with scaled
experimental data of Smid and Novosad and (*) that of Brockbank et al 
(averaged over three piles). 
Upper and lower curves denote normal and shear stresses. 
The data is used to calculate the total weight of the pile which is then 
used as a scale factor for stresses. The horizontal coordinate is scaled 
by the pile radius.
\label{data}}
\ec
\efig

We believe that the experimental data strongly
suggest an {\sc fpa}-like model (corresponding to $c_0^2\simeq 1$,
Eq.\ref{fpa}),
although they do not prove that the principal axes of a material element are
necessarily fixed. Of greater physical significance is the idea that the
characteristic rays of the wave equation could be fixed at burial.
(Below we will connect this with the idea of stress paths.) This is the
content of a more general interpretation (the ``oriented stress linearity"
model of \cite{FPA}) in which $\mu$ (or $c_0^2$) is a parameter characterising 
the local anisotropy of the granular packing.  
This is fixed by construction history, in a manner as yet unspecified; 
but for a point-source pile values close to the {\sc fpa}
limit are suggested by the data. In contrast, for a pile constructed by
sieving, we
might expect the local reflection symmetry to be unbroken ($\mu =0$), leading
instead to a smooth maximum in the pressure, as predicted from Eq.\ref{bcc}.
Experiments on the construction-history dependence of the stress profile would
therefore be welcome.

Note that the idea that some property of the medium, represented
by a constitutive equation among stresses, is ``fixed at burial" is an
important simplifying assumption of our approach to sandpile modelling. This
has been called the ``perfect memory" assumption \cite{FPA}.

\subsection{Wedges: Surface Avalanches or Deep Yield?}

Savage \cite{Savage} has pointed out that some classical data on wedges, as
opposed to cones, are apparently at odds with these ideas. Such wedges are of 
triangular cross-section but very long in the third direction, {\it i.e.}, 
quasi-two dimensional.  According to classical reports \cite{Hummel,Lee}, 
the stress distribution beneath such wedges shows little dependence on the 
construction history, but a very strong dependence on whether the base 
supporting the wedge is allowed to sag. Without basal sag, almost no pressure 
dip is reported \cite{Savage}.

At first sight, then, the pressure dip in conical piles might also be 
attributable to basal sag. We strongly believe that this viewpoint is 
untenable, especially in the light of the data of Brockbank et al 
\cite{Huntley} (see figure \ref{data}) in which the
measurement system leads to  discernable localized indentation, rather than
a slight curvature (sagging) of the base under regions of high pressure.
Moreover, the classical data on wedges is somewhat scant, of dubious
accuracy, and in most cases does not specify the construction history of 
the wedge in any detail.
We would therefore encourage renewed experimental investigation of wedges
of sand.

There is, in any case, an important difference in the physics of wedges and
cones.
This concerns the geometry of the ``plastic" region (that in which the
Mohr-Coulomb inequality is saturated) near the surface of the pile. 
All our models, including the {\sc fpa} limit \cite{FPA}, predict that this 
is infinitely narrow in the three dimensional conical pile, but that for 
a wedge it extends through a large outer zone
(exterior to the two characteristic rays passing through the apex). 
In the first case, the perfect memory assumption appears self-consistent: 
the presence of a thin yield layer at the surface suggests that the pile 
grows by surface avalanches which
do not disturb its internal structure too much. In contrast, the status of the
perfect memory assumption is, for a wedge, far less clear. Since a broad zone of
marginal instability exists beneath the surface, the surface flow could
cause ``deep yield" events \cite{Evesquedy,Evesque} which would disrupt the 
internal structure of the pile \cite{usunpub}. This could lead to a local 
isotropisation of the granular texture and cause values of $\mu$ in 
(\ref{osl}) closer to those of the symmetric propagation model (Eq.\ref{bcc}) 
than the {\sc fpa} model (Eq.\ref{fpa}).  Much more
detailed experimental data is certainly needed, however, before these ideas
can be put to the test.

\section{The R\^ole of Local Inhomogeneities}

\subsection{A stochastic wave equation}

Provided that local conservation laws (those arising from
mechanical equilibrium)  are obeyed, many local rules for force transmission are
{\it a priori} compatible with the existence of contacts among rigid particles
\cite{Eloy,Socolar}.  Therefore, even if  there is a definite mean relationship
among stresses at the meso-scale (as models like {\sc fpa} assume), one can
expect randomness in the local transmission coefficients. 
The simplest model for this and other sources of randomness is
to introduce a randomly varying `speed of light'
$c_0$. This could describe the fact that, for example, the parameter $p$
in the model of figure \ref{threeleg} can vary from grain to grain.

This suggests the following stochastic wave equation for
stress propagation in two dimensions:
\be
\partial_{tt} \szz = \partial_{xx} \left[c_0^2(1+v(x,t))\ \szz\right]\label{RW}
\ee
where $v(x,t)$ is a random noise. We have studied this equation in great
details \cite{noise} using perturbation theory, and find that for weak disorder,
the average response function now has two peaks of finite (diffusive)
width (rather than two $\delta$ peaks in the zero disorder case), and that the
`speed of light' is renormalized to a lower value $c_R$. 
More interestingly,  the unaveraged response function takes negative 
(and rather large) values. This may be of crucial importance since it 
suggests a fundamental instability of granular matter to external 
perturbations. Suppose indeed that as a
result of a distant perturbation, a certain grain receives a negative
(upward) force larger than the pre-existing downward vertical pressure. 
This grain will then move and a local rearrangement of contacts will occur. 
If stability is to be recovered,
this rearrangement must induce a variation of
$c_0(x,t)$ so as to reduce the cause of the instability. Thus, the stochastic wave
equation implicitly demands rules similar to those introduced in \cite{CB} to
describe extreme sensitivity to external perturbations in silos. The present
model, which is purely static, does not say what happens when a local
rearrangement occurs, but certainly suggests that small perturbations will 
induce such rearrangements. One can show that typically a perturbation of 
order the weight of one grain is enough to oblige rearrangements somewhere 
else in the pile \cite{noise}.
This would cease to be true if large enough overload was applied to ensure that
that all grains are subject to a vertical compression greatly in excess of their
weight. But it is not clear that such an overload is ever really possible:
even at the bottom of a deep pile, a finite fraction of grains may be 
effectively non-loadbearing \cite{Nagel}.

We have also studied the weight-weight correlation function, in the
geometry of a
flat layer with a random overload at the top, neglecting gravity. We
find \cite{noise} that, as a function of (horizontal) distance, the
correlation has {\it two} peaks.  The first one is of course at 
separation $\Delta x=0$, while the second is at
$\Delta x=2 c_R H$, which simply means that two points at the bottom of the
packing
connected by `light rays' to the same point on the top, share information
about the
overload. This result is of importance since the shape of this correlation
function
clearly differs from the corresponding one in the scalar model, which is a
single peak at $\Delta x=0$. 
It also differs from that pertaining to a simple elastic medium, where
the weight-weight correlation, in this geometry, decays only very slowly:
correlations extend to the scale of the system size itself. Measuring
carefully the
averaged correlation function of a granular system under an overload could then
confirm (or disprove) the presence of a ray-like  propagation. A stress
correlation
function was recently measured in \cite{Nagel} and found to be featureless, but
measurements extended only to very short lengthscales: $\Delta x \leq 5 a$, as
compared to the height of the pile $H
\simeq 100 a$. We thus expect the features of the correlation function to
show up on
much larger scales ($\sim 2 c_R H$) than those measured so far.

\subsection{Stress histogram}
\label{numerics}

We have seen above
that within a scalar approach, an exponential-like distribution (possibly
of the type $\exp -w^\beta$, with $\beta \geq 1$) is expected 
\cite{Liu,Copper}. One can ask whether this exponential
distribution survives within a tensorial description.
So far, we only have partial numerical results, based on a direct
simulation of the three-leg model introduced above, with a random $p$ chosen
between $0$ and $p_M$. This scheme is thus very close in spirit to the
$q$-model.
However, as emphasized above, the local vertical forces are not everywhere
positive; one should thus introduce an extra rule to cope with this instability.
Several possibilities come to mind, but we have not yet explored them (see
however
\cite{Eloy,Socolar}). Nevertheless, the large force region, which is
presumably not
sensitive to the presence of negative forces, behaves much in the same way as in
the scalar $q$-model. In particular, the tail of the distribution decays as
$\exp
-w^\beta$, with $\beta \simeq 1$ when $p_M=1$, and with $\beta > 1$ when 
$p_M < 1$.  
More work is needed to understand the physical implications of the presence
of negative forces and any relation this may have to the static avalanche
phenomenon \cite{CB}. However, the above results show that the tail of the
force distribution is only exponential in a `strong disorder' limit, 
where local `arching' (i.e. one grain entirely bearing on a single downward 
neighbour) has a nonzero probability.

\section{Modelling Strategies for Static Granular Media}
We now compare our approach (as outlined above) to the problem of
sandpile modelling with previous ones, and address various criticisms of it
that have recently been made. Specifically we are interested in calculating 
the stress
distribution at the base of a pile constructed on a {\it rough, rigid}
support. For
clarity of the discussion, we again limit the mathematics to two dimensions,
although we emphasize again that our work, particularly that on the 
{\sc fpa} model, was developed in the context of three dimensional conical 
piles.

\subsection{Local Rules for Force Transfer}
Above we have described an approach to the modelling of stress propagation
in granular media based on local rules for the transfer of forces between
grains.
Broadly speaking, to leading order in a gradient expansion
\cite{usunpub,noise} the
closure (\ref{osl}) exhausts the possibilities for local models in which
the forces
passed from a grain to its neighbours in the layer below involve a
linear decomposition of the ``incident force" $(F_x,F_t)$,
which is taken as the vector sum of forces acting on the grain from those
in the layer
above. Somewhat similar considerations underlie the
so-called ``clastic" theories of ``discontinua" introduced  by Trollope
\cite{Trollope}. Indeed, some authors have confused the two models, and our
approach has been criticized for ``reinventing theories... that are 
well-known in another disciplinary area" \cite{Savage}.

However, Trollope's model, though linear, does not
lead to Eq.\ref{osl}. Its distinguishing feature is instead
that {\it the vector sum of the incident forces on a grain is not taken} before
applying a rule to determine the outgoing forces from that grain.  The outgoing
forces instead depend {\it separately} on each of the incident force
contributions.
This feature is, in our view, strongly unphysical: if a grain is subjected
to two small extra forces, whose vector sum is
vertical, from its neighbours in the layer above,  the force increments
exerted on the grains below should be equivalent to a small increase in its
weight.
Within Trollope's rules, this is not the case  \cite{trolfoot,usunpub}.
It is therefore
disingenuous to criticize our approach  on the grounds that  `Trollope's
clastic or discontinua model, that was rejected by Wittmer et al. on the
grounds of
being ``unphysical", actually contains the {\sc fpa} solution'
\cite{Savage}. For, although Trollope's model can be tuned to give the same
stress pattern as the {\sc fpa} model in a symmetric two dimensional pile, 
these two models have distinctly different physical content and give 
differing predictions in other geometries.

\subsection{Hyperbolic Continuum Models}
Although motivated in part by simple packing models (e.g. figure
\ref{threeleg}),
our approach does not require such a specific microscopic interpretation and 
can instead by formulated in purely continuum mechanical terms \cite{footcont}.
This is done by postulating a constitutive relations among stresses such as
(\ref{osl}). (The latter includes, for special values of $\mu$, both
(\ref{bcc}) and (\ref{fpa}) which describe respectively the case of
symmetric stress
propagation and the {\sc fpa} model.)
We have noted already the existence of an alternative constitutive equation
among
stresses, corresponding to the Mohr-Coulomb rigid-plastic model (or {\sc
ife} model)
which is a widely studied continuum theory of granular media. In fact this reads
(in two dimensions):
\be
\sigma_{xx} = \sigma_{zz}\,{1\over \cos^2\phi}\left[
\sin^2\phi+1 \pm
2\sin\phi\,\,\sqrt{1-(\cot\phi\;\;\sigma_{zx}/\sigma_{zz})^2}\right]\label{ife}
\ee
All these approaches lead to hyperbolic equations for stress propagation.

A reasonable question \cite{Savage} is to ask why we are not satisfied with
the {\sc ife} closure (\ref{ife}). The main cause is that we can see no 
physical reason for it to be true. Indeed, even supporters of this 
rigid-plastic model do not usually propose it as an accurate description of 
sandpile behaviour; it is more often viewed as a way of generating certain 
``limit-state" solutions.
In the simplest geometries these solutions correspond to taking
the $-$ or $+$ sign in (\ref{ife}); these are often referred to as 
``active" and ``passive" limit-states, respectively \cite{footpass}. 
It is easily established \cite{FPA,usunpub} that for a sandpile at its 
repose angle, only one solution of the resulting equations
exists in which the sign choice is everywhere the same. 
This is the active solution, and it shows a hump, not a dip, in the 
vertical normal stress beneath the apex.
Savage, however, draws attention to a ``passive" solution,
having a pronounced dip beneath the apex \cite{Savage}. This solution actually
contains a pair of matching planes between an inner region where the
positive root of (\ref{ife}) is taken, and an outer region where the 
negative is chosen.

\bfig[hbt]
\bc
\epsfysize=6cm
\epsfbox{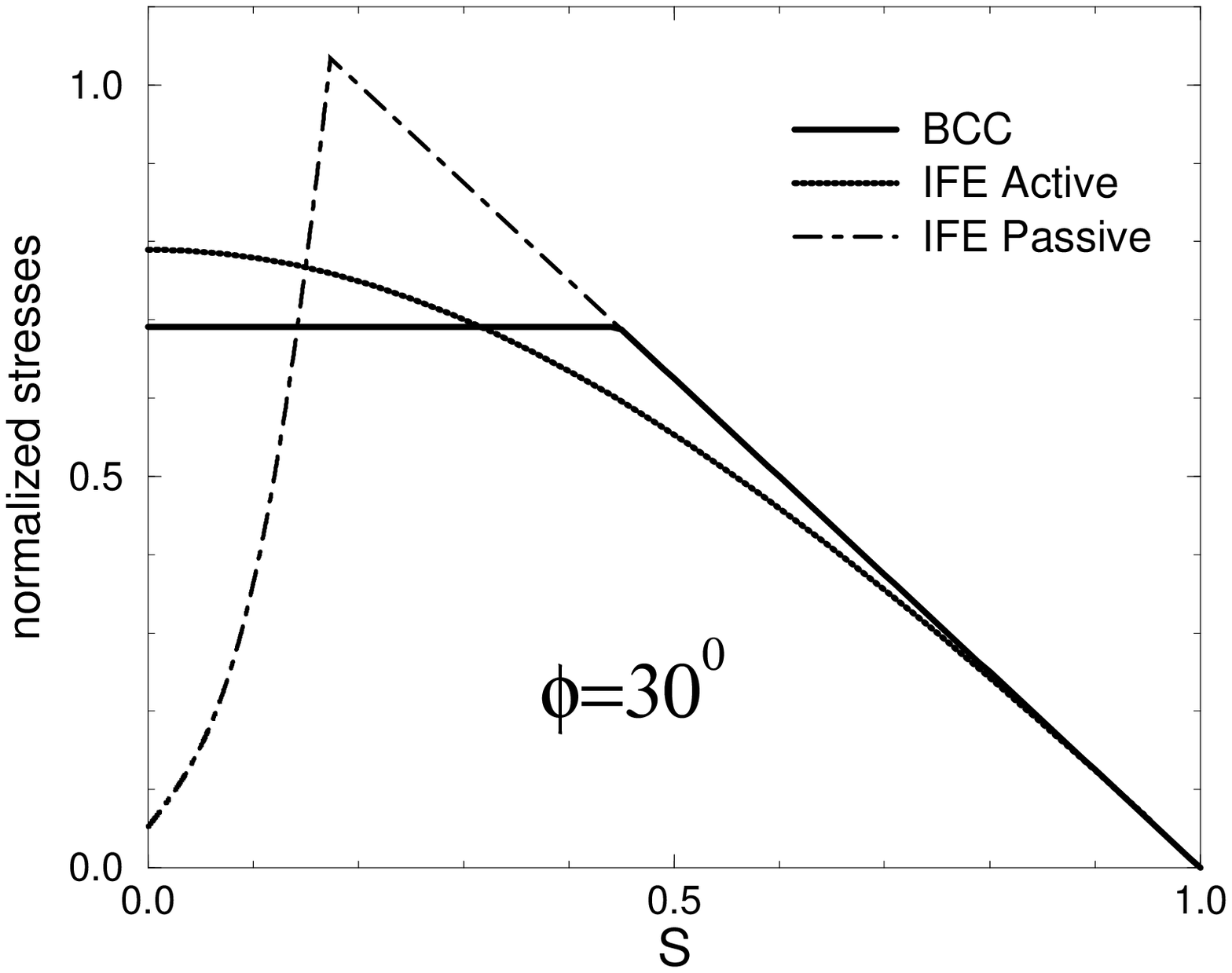}
\caption{Vertical normal stress found from Eq.\protect\ref{bcc} (the  {\sc bcc}
model, \protect\cite{BCC}), for a pile at angle of repose $\phi=30$
degrees, compared to the active and passive {\sc ife} solutions.
(The {\sc ife} solutions are obtained by shooting from the midplane for
$P=(\sigma_{tt}+\sigma_{xx})/2$ and the polar angle $\theta$ as functions
of the direction of the principal axis $\Psi$.)
Note that active and passive {\sc ife} solutions do not bound the stress,
either in the model of Eq.\protect\ref{bcc} or in the elastoplastic model of
\protect\cite{Cantelaube}, which, for a certain parameter choice, yields identical
results to Eq.\protect\ref{bcc} (see \protect\cite{Cantelaube}).
\label{bounds} }
\ec
\efig
In principle this does not exhaust the repertoire of {\sc ife} solutions for
the sandpile: there should exist others, with larger numbers of matching planes
between segments of alternating sign \cite{usunpub}.  In any case, the
predictive power of the rigid-plastic approach largely depends on a belief 
that the limit-state solutions can be ``generally regarded as bounds between 
which other states can exist, {\it i.e.}, when the material is behaving in 
an elastic or elastoplastic manner" \cite{Savage}. Unfortunately, this belief 
is unfounded: counterexamples exist (figure \ref{bounds}), even among elastoplastic models of the 
simplest kind \cite{usunpub}.
Therefore the so-called limit-states should be regarded merely as ``rule of
thumb"
estimates; these may be useful for engineering purposes, but do not shed
much light on the physics of stress propagation in granular matter.

\subsection{Elastoplastic Models}
All the models considered above make no mention of strain variables. A partial
justification for this was given in Refs.\cite{WCCB,FPA}, namely that the
experimental data obey radial stress-field (RSF) scaling, which implies
that there is no characteristic length-scale. Since elastic deformation 
under gravity introduces such a length-scale (a ``sagging length") the 
observation of RSF scaling to experimental accuracy in most but not all the 
data \cite{Smid,Huntley,Savage} suggests that elastic deformation is not 
significant.

This does not imply that an elastic or elastoplastic description of sandpiles 
is impossible; but it shows that in any such description, the limit of a large
elastic modulus appears to be the relevant one. This limit yields equations, 
in the bulk of the medium, for which strain variables cancel out; 
and this fact is usually exploited in elastoplastic calculations 
(see, e.g., \cite{Cantelaube}). Note that it is tempting, but entirely wrong, 
to assume that strain variables on the boundary of the medium also cancel in 
this limit (see below).

In fact, as correctly noted by Savage \cite{Savage} results similar
to those of the {\sc fpa} model and the related models described above can,
in two dimension (only), be obtained within such an elastoplastic modelling 
approach. Typically, an inner, linear elastic region is matched, by hand, 
onto an outer plastic one. An example of this procedure was described 
recently by Cantelaube and Goddard \cite{Cantelaube} whose approach is 
similar to earlier work by Samsioe \cite{Samsioe}. 
This analysis can be made to give
mathematically identical results to those found with some, if not all,
values of $\mu$ in Eq.\ref{osl}.

As they stand, however, such elastoplastic analyses are devoid of physical
meaning, for the following reason. Recall that the aim of the exercise is to
calculate the forces measured at the base of a pile of sand. Recall also the
well-known theorem that to find the equilibrium state of an elastic body, one
must specify either the surface force field or the displacement field at
all points on its boundary \cite{landau}. Accordingly, it is meaningless to
``calculate" the
forces at the base of an elastoplastic pile without specifying a boundary
condition at the bottom surface of any elastic zones present. To specify as
boundary conditions the forces themselves there is mathematically
legitimate, but
cannot really be described as a ``solution" to the sandpile problem. On the
other
hand, to specify the displacements is (by this stage) formally impossible,
since the
relevant variables have already been eliminated by taking the limit of a
high elastic modulus.

The only way to circumvent this difficulty is to specify the
displacement field at the base first, and then take the limit of a high modulus
afterwards. To obtain finite forces, the displacements must be allowed to
tend to zero but, crucially, the results depend on {\it how this
limit is taken}. This is illustrated in figure \ref{indeterminate}.
The challenge to elastoplastic modellers is then to decide what
(infinitesimal) displacement field should be chosen at the base. For a genuine
elastoplastic body, which is placed on a rough rigid support, this displacement
field depends on the precise manner in which the body was brought into its
state of rest. For a sandpile constructed by (say) pouring sand grainwise 
from a point source the problem seems much less well-posed since (despite
assertions to the contrary \cite{Savage}) there is no obvious physical
definition of the displacement field in such a pile
\cite{footstrain}.
\bfig[hbt]
\bc
\epsfysize=4.5cm
\epsfbox{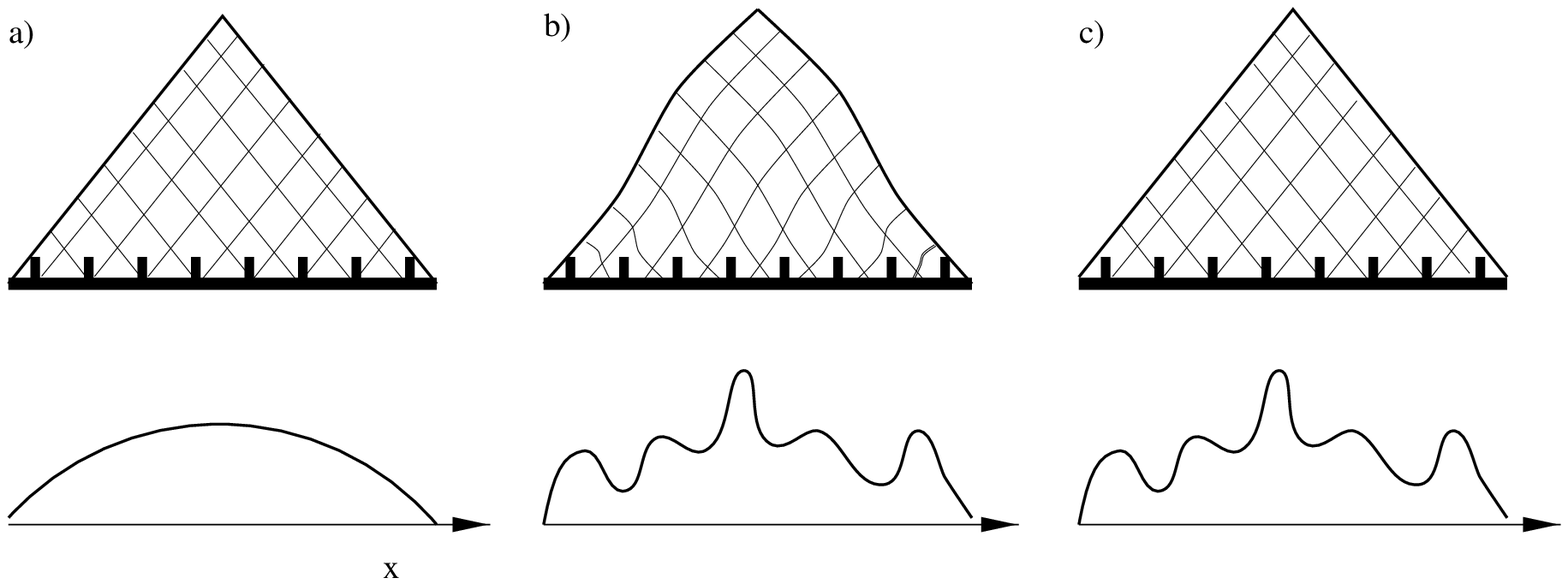}
\caption{Consider the static equilibrium of an elastic
body of finite modulus resting on a completely rough
surface.
Starting from any initial configuration, another can be generated
by pulling and pushing parts of the body horizontally across the
base ({\em i.e.}, changing the displacements there); if this is rough,
the new state will still be in static
equilibrium. This will generate a stress distribution, across the
supporting surface and within the pile, that differs from the
original one. If the limit of a large modulus is now taken (at
fixed stress), this procedure allows one to generate arbitrary
differences in the stress distribution while generating
neither finite distortions in the shape of the body, nor any
forces at its free surface. An exactly analagous ``elastic indeterminacy"
exists in any
simple elastoplastic theory of sandpiles, where an elastic zone, in
contact with part of the base, is attached at matching surfaces
to a plastic zone.
\label{indeterminate}}
\ec
\efig

\subsection{Elastic Indeterminacy}\label{hooky}
The circumstances just described apply to
any model of a sandpile in which the elliptic equations of elasticity are
invoked in
all or part of the medium. We have called this the problem of ``elastic
indeterminacy" \cite{usunpub}.
Several responses to this challenge are possible. One is
to assume that the elastoplastic description is essentially sound, and
seek some physical procedure whereby the displacement field at the support, or
some equivalent information, is determined \cite{tom}. This is certainly worth
pursuing, but we suspect that any description which does not
explicitly consider the construction history of the pile is very unlikely to be
successful.

A second response, which appears to be that of Evesque
\cite{Evesque} is to conclude that
the experimental results {\it are and must be indeterminate}. Put differently,
Evesque argues that the external forces acting on the base of a pile can
in fact be varied at will by the experimentalist. Certainly, if sandpiles obey
Hooke's law, he must be right: for an elastic body, the experimentalist is free
to prod about at the base of a pile in such a way as to change arbitrarily
the forces acting there. 
(Moreover, in the large modulus limit, only infinitesimal proddings
are required.)

In contrast to this, experimental reports clearly suggest
\cite{Smid,Huntley} that the forces on the base can be measured more or less
reproducibly, and (though subject to statistical fluctuations) do not vary
too much from one pile to another.  Moreover, the experimental
data for the forces show RSF scaling at the base; this is confirmed by
comparing data from piles of different heights. Within an elastoplastic
framework,
this scaling should be violated by the (arbitrary) boundary forces at the
base and
hence can only be expected in the upper extremity of the pile
\cite{Cantelaube,usunpub}.

All this suggests a third response to the
challenge of elastic indeterminacy. This is to argue that Hooke's law has
very little
relevance to the mechanics of sandpiles and that models based on local force
propagation rules among grains  are far closer to the real physics of the
problem.
The models we have developed along these lines give hyperbolic equations for
stress propagation, and in doing so contradict Hooke's law in a fundamental
way. This
applies even in their incremental response to small added loads. Whether
this is a
drawback or an advantage depends on one's  view of the physics. Certainly,
if one believes
that any granular assembly must behave elastically under sufficiently
small incremental loads, then models such as ours  can only describe the
behaviour
beyond some finite threshold \cite{tom}. (We are as yet unconvinced of
whether this
threshold is finite; for example it might vanish in the high modulus
limit.)  In any
case, in calculating the response to gravity itself we believe that any such
threshold is easily exceeded throughout the pile, and hence that our
approach has
far more to offer than models invoking elastic deformations from some
hypothetical
unstrained ({\it i.e.}, zero gravity) state. On the other hand, the
existence of a
threshold might make it somewhat harder to detect experimentally the
response and correlation functions
\cite{BCC,FPA,noise} which are, as described above, strong signatures of
hyperbolic stress propagation laws.

The varying perspectives laid out above are not
necessarily completely contradictory, in that the global
(`coarse-grained')  features of the stress pattern could be governed by
determinate equations but the details not.  As well as being subject to
elastic indeterminacy, the latter could be affected by randomness in the grain
packings, which may be exquisitely sensitive to temperature and other
poorly-controlled parameters \cite{CB}.

\section{Boundary Conditions in Hyperbolic Models}\label{charz}
We now consider more carefully the role of boundary conditions in our approach.
In contrast to the elliptic equations of elasticity, the hyperbolic
equations arising from constitutive relations among stresses  admit definite
solutions for the stresses at the base of a freestanding pile. By the
same token, if one tries to apply all the boundary conditions appropriate to an
elastic body (e.g. specifying the surface forces acting over the entire
surface),
then in general no solution will exist of the hyperbolic equations that
pertain to any particular choice of constitutive relation. We
discuss these two aspects in turn.

\subsection{Determinacy of the Sandpile}
The procedure for a sandpile is of course to specify zero-force boundary
conditions
at the free (upper) surfaces. Within our description, the response
arising from a localized body force (a ``source term" in the equations)
propagates
downward along two characteristics passing through the source. In models obeying
(\ref{osl})  these characteristics are, in addition, straight lines. The
force on
the base is found simply by summing the contributions from all the body forces;
this is a fully determinate procedure for any closed set of hyperbolic equations
\cite{BCC,WCCB,FPA}.
Note that in principle, one
could envisage propagation also along the ``backward" characteristics (see
figure
\ref{pathfig}(a)). This is forbidden since these cut the free surface; any such
propagation can only arise in the presence of a nonzero surface force, in
violation
of the boundary conditions. Therefore the fact that propagation occurs only
along
downward characteristics is not related to the fact that gravity acts
downward; it
arises because we know what forces act at the free surface (the forces there are
zero). Suppose we had instead an inverse problem: a pile or bed with some
unspecified
overload at the upper surface, for which the forces acting at the base had been
measured. In this case, the information from the known forces could be
propagated
along the {\it upward} characteristics to find the unknown overload.
\bfig[hbt]
\bc
\epsfysize=6cm
\epsfbox{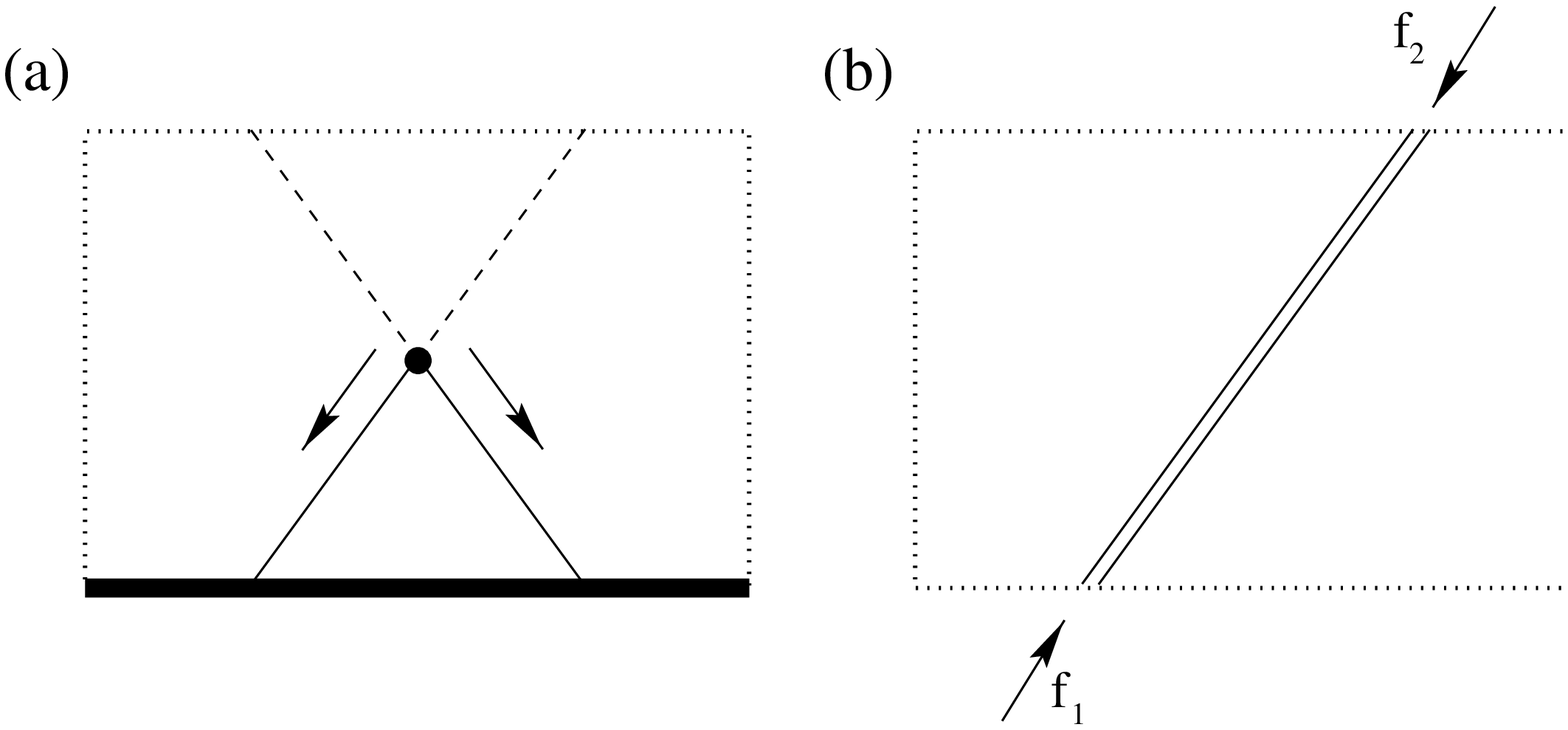}
\caption{(a) The response to a localized force is found by resolving it along
characteristics through the point of application, propagating along those
which do
not cut a surface on which the relevant force component is specified.
For a pile under gravity, propagation is only along the downward rays.
(b) Admissible boundary conditions cannot specify separately the force
component at both ends of the same characteristic. If these forces
are unbalanced (after allowing for body forces), static equilibrium is
lost.\label{pathfig}}
\ec
\efig

\subsection{Stress Paths}
Informally speaking, the hyperbolic problem is determined once half of the
boundary  forces are specified. More precisely (figure
\ref{pathfig}(b)) one is required to specify the surface force
tangential to each characteristic ray, at one end and {\it one end only}. The
corresponding force acting at the other end is obliged to balance it,
allowing for any body forces acting tangentially along the ray. If it does
not do so,
then within our modelling approach, the material ceases to be in static
equilibrium.
This is no different from the corresponding statement for a fluid or liquid
crystal;
if boundary conditions are applied that violate the conditions for static
equilibrium, some sort of motion results. Unlike a fluid, however, for a
granular medium we expect such motion to be in the form of a finite 
rearrangement rather than a steady flow. Such a rearrangement will change 
the microtexture of the material, and thereby {\it alter the constitutive 
relation among stresses}. We
expect it to do so in such a way that the new constitutive relation is
compatible with the imposed forces.

Although simplified, we believe that this picture correctly captures some of the
essential physics of stress paths. Such paths are load-bearing structures
within the
contact network and, in the simplest approximation of straight, unbranched
paths,
these must have the property described above (the forces on two ends
of a path must balance). Stress paths should therefore be identified (on
the average) with the characteristics of our hyperbolic equations. 
The {\it mean orientation} of the stress paths is then reflected in a 
constitutive equation such as (\ref{bcc}) or (\ref{fpa}).

The physics of our modelling approach is thus to assume that the mean
orientation of stress paths, in each element the material, is fixed at
burial. (This does not necessarily require that the individual paths are 
themselves fixed.) This was called earlier the ``perfect memory assumption".  
We think it reasonable to assume that the stress paths will not change their 
average orientations so long as they are able to support the applied load. 
But if a load is applied which they cannot support, rearrangement is 
inevitable \cite{Evesque}. This
causes some part of the pile, at least, to adopt a new microtexture and
thereby a new
constitutive relation. In other words, loadings of this kind cause the
construction history of the pile to change.

\section{Conclusion: Sandpiles as Fragile Matter}
Within our modelling approach, a granular
assembly is able to support some, but not all, of the surface loads that
would be supportable by an elastic continuum.
Such models may therefore provide an
interesting paradigm for the behaviour of ``fragile matter", a
concept which may be useful in other systems where certain combinations of
applied
forces, even if small, are enough to force irreversible reconstruction of the
material. (Such systems could include a number of disordered soft solid
materials such as defect textures in liquid crystals.)

In the present context, fragility arises from the the requirement of
tangential force balance along stress paths. If this is violated at the 
boundary, even infinitesimally, then internal rearrangement must occur, 
causing new stress paths to form, so as to support the load \cite{ifesen}.  
Obviously, this might be rather too
simple a picture -- for example, branching of stress paths is ignored. 
Thus it remains possible that Hookean behaviour is recovered for sufficiently 
small perturbing forces \cite{tom,Evesque}. However, for practical purposes we
believe our approach, by capturing at least some of the physics of stress 
paths, may have rather more to offer than ideas based on the physics of 
conventional (homogenous, isotropic) elastic, elastoplastic or rigid-plastic 
media. Phenomena that seem to be addressable in such terms include that of 
arching and, in its various aspects, the role of noise.

Clearly, there is much scope for developing these models further. In
particular it would be very useful to have an understanding of the crossover, 
if one indeed exists, between fragile and elastic regimes: 
the latter should ultimately be restored in a
sufficiently large pile (beyond the ``sagging" length). Equally important is 
to confront these and other modelling approaches with much more demanding
experimental tests, of which several were suggested above.

\section{Acknowledgements} We are grateful to P. Evesque, J. Goddard, J.
Jenkins, S. Savage and F. Radjai for discussions, and to H. Herrmann for
facilitating  some of these during the present School. MEC acknowledges the
hospitality of ITP Santa Barbara where some of these ideas were developed during 
discussions with S. Edwards, D. Levine, S.  Nagel, C. Thornton, T. Witten 
and other participants of the ``Jamming and Rheology"
programme. This research was funded in part by EPSRC (UK) Grants GR/K56223 and
GR/K76733.


\newpage

\begin{thebibliography}{99}

\bibitem{Smid}
J. Smid and J. Novosad,
Proc. of 1981 Powtech Conference, Ind. Chem. Eng. Symp. {\bf 63}, D3V 1-12
(1981).

\bibitem{Huntley}
R. Brockbank, J.M. Huntley, and R.C. Ball,
J. Phys. II (France), {\bf 7}, 1521-1532 (1997).

\bibitem{WCCB}
J.P. Wittmer, M.E. Cates, P. Claudin and J.-P. Bouchaud,
Nature (London) {\bf 382}, 336 (1996).

\bibitem{Brown}
R.L. Brown and J.C. Richard,
`Principles of Powder Mechanics' (Pergamon, New York, 1966).

\bibitem{Clement}
L. Vanel, E. Cl\'ement, J. Lanuza and J. Duran,
preprint submitted to Phys. Rev. Lett.; and this volume.

\bibitem{Radjai}
F. Radjai, M. Jean, J.-J. Moreau and S. Roux,
Phys. Rev. Lett. {\bf 77} 274 (1996);
F. Radjai, D.E. Wolf, M. Jean and J.-J. Moreau,
{\it Binomal character of stress transmission in granular packings}, preprint;
and this volume.

\bibitem{Baxter}
G.W. Baxter, in {\it Powders and Grains 97}, Behringer and Jenkins eds.,
Balkema, Rotterdam (1997).

\bibitem{Dantuetc}
see e.g. P. Dantu, Proc. of the fourth Int. Conf. On Soil Mech. and Found. Eng. (London 1957), {\bf 1} 144, Annales des Ponts et Chauss\'ees, {\bf IV}, 193 (1967),
T. Travers et al.,  J. Phys. France, {\bf 49} 939 (1988).

\bibitem{Liu}
C.-H. Liu, S.R. Nagel, D.A. Scheeter, S.N. Coppersmith, S. Majumdar, O.
Narayan and T.A. Witten, Science {\bf 269}, 513 (1995).

\bibitem{Nagel}
D.M. Mueth, H.M. Jaeger and S.R. Nagel,
preprint

\bibitem{Copper}
S.N. Coppersmith, C.-h. Liu, S. Majumdar, O. Narayan and T.A. Witten,
Phys. Rev. {\bf E 53}, 4673 (1996).

\bibitem{BCC}
J.-P. Bouchaud, M.E. Cates and P. Claudin,
J. Phys. I (France) {\bf 5}, 639 (1995).

\bibitem{FPA}
J.P. Wittmer, M.E. Cates and P. Claudin,
J. Phys. I (France) {\bf 7}, 39 (1997).

\bibitem{Gudehus}
See, e.g., G. Gudehus in {\it Powders and Grains 97}, 
Behringer and Jenkins eds., Balkema, Rotterdam (1997), p. 169-183.

\bibitem{Savage}
S.B. Savage in {\it Powders and Grains 97}, Behringer and Jenkins eds.,
Balkema, Rotterdam (1997), p. 185-194;
see also New Scientist, {\bf 2083}, p.28 (1997).

\bibitem{Cantelaube}
F. Cantelaube and J.D. Goddard,
in {\it Powders and Grains 97},
Behringer and Jenkins eds., Balkema, Rotterdam (1997), p231-234.

\bibitem{Evesque}
P. Evesque and S. Boufellouh, in {\it Powders and Grains 97},
Behringer and Jenkins eds., Balkema, Rotterdam (1997) p.295-298; P. Evesque, J.
Physique I, {\bf 7}, 1305-1308 (1997); P. Evesque, private communication.

\bibitem{Edwards}
S.F. Edwards and R.B. Oakeshott,
Physica {\bf D 38}, 88-93 (1989);
see also S.F. Edwards and C.C. Mounfield,
Physica {\bf A 226}, 1,12,25 (1996).

\bibitem{CB}
P. Claudin and J.-P. Bouchaud,
Phys. Rev. Lett. {\bf 78}, 231 (1997); and this volume.

\bibitem{Janssen}
H.A. Janssen, {\em Z. Vert. Dt. Ing.} {\bf 39}, 1045 (1895);
see also R.M. Nedderman,
{\em Statics and Kinematics of Granular Materials},
Cambridge University Press (1992).

\bibitem{usunpub}
M.E. Cates, J.P. Wittmer, J.-P. Bouchaud and P. Claudin,
manuscripts in preparation.

\bibitem{footpod}
For a more general geometry (without axial symmetry) at least
two further closure relations are required. Possibilities include not only
isotropic, wavelike propagation (in $2+1$ dimensions), but also propagation
along three characteristic rays in the form of a tripod. The latter is perhaps
the most natural extension of the {\sc fpa} hypothesis to arbitrary geometries.

\bibitem{Hummel}
F.H. Hummel and E.J. Finnan, Proc. Inst, Civil Eng.
{\bf 212}, 369-392 (1920).

\bibitem{Lee}
I.F. Lee and J.R. Herington, Proc. 1st Aust.-N.Z. Conf. Geomech.,
{\bf 1}, 291-297 (1971).

\bibitem{Evesquedy}
P. Evesque,  Phys. Rev. {\bf A 43}, 2720 (1991);
P. Evesque, D. Fargeix, P. Habib, M. P. Luong and P. Porion,
Phys, Rev. {\bf E 47}, 2326-2332 (1993).

\bibitem{Eloy}
C. Eloy and  E. Cl\'ement,
to appear in J. Phys. I (France) (1997).

\bibitem{Socolar}
J. Socolar, preprint, cond-mat/9710089.

\bibitem{noise}
P. Claudin, J.-P. Bouchaud, M.E. Cates and J. Wittmer,
preprint, cond-mat/9710100, submitted to Phys. Rev. {\bf E}. 

\bibitem{Trollope}
D.H. Trollope, in {\em Rock Mechanics in
Engineering Practice}, pp. 275-320, K.G. Stagg and O.C. Zienkiewicz,
(Eds.), (Wiley, New York, 1968). D.H. Trollope and B.C.Burman,
G\'eotechnique {\bf 30}, 137-157 (1980). 

\bibitem{trolfoot}
For nonzero values of his arching parameter $k$. For $k=0$,
the continuum limit of Trollope's model coincides with \ref{bcc}. However, this
is not the case discussed by Savage \cite{Savage}.

\bibitem{footcont}
It would be a mistake to suggest that our modelling
strategy is fundamentally at odds with continuum-mechanical principles
\cite{Cantelaube} (or even with the laws of newtonian mechanics themselves
\cite{Evesque}). Such remarks seem to be based on the idea that Hooke's law is
implicit in any continuum (or even newtonian) description. This is untrue,
as the existence of successful continuum theories of fluids and liquid crystals
shows. We return below (Section \ref{hooky}) to the issue of whether Hooke's law can
usefully be applied to granular media under gravity.

\bibitem{footpass}
Our definitions of ``active" and ``passive" are not quite 
the same as those used elsewhere in the literature, in which the terms refer 
to global properties of the solutions rather than the choice of a local
constitutive equation. According to the latter, Savage's identification of 
the (+/-) solution
containing a matching plane as everywhere passive is correct, although in our
terminology it a solution with an active outer and passive inner region.

\bibitem{Samsioe}
A.F. Samsioe, G\'eotechnique {\bf 5}, 200-223 (1955).

\bibitem{landau}
See, e.g., L.D. Landau and E.M. Lifshitz,
{\em Theory of Elasticity}, 3rd Edn., Pergamon, Oxford
1986, or G. E. Mase, {\em Continuum Mechanics}, McGraw Hill, NY
1970.

\bibitem{footstrain}
Since a sandpile would not exist at all in the absence of
gravity, it is not clear whether one can as usual define displacements
relative to a reference state in which no body or surface forces act.

\bibitem{tom}
We are grateful to Tom Witten and others for discussions on this point.

\bibitem{ifesen}
There is a sense in which this can be viewed as ``incipient failure everywhere",
except that the failure in question is not Mohr-Coulomb, but instead connected
with the failure of stress paths under imbalanced tangential loads.

\end{thebibliography}
\end{document}